\documentclass[twocolumn,11pt]{wseas}

\usepackage{abstract, fancyhdr,epsfig,rotating}
\DeclareFixedFont{\Tifont}{T1}{ptm}{b}{it}{16pt}
\renewcommand{\thefootnote}{\fnsymbol{footnote}}

\title{GuiLiner: A Configurable and Extensible Graphical User Interface for 
Scientific Analysis and Simulation Software}
\author{Nicholas C. Manoukis\footnotemark[2] 
\and Eric C. Anderson\footnotemark[4] \and
\\ \footnotemark[2] Section of Vector Biology, Laboratory of Malaria and Vector Research \\
National Institute of Allergy and Infectious Diseases, National Institutes of Health \\
12735 Twinbrook Parkway, Bethesda, MD 20892 USA\\ 
\vspace{2mm} \\
\footnotemark[4] Fisheries Ecology Division, Southwest Fisheries Science Center
\\ National Oceanic and Atmospheric Administration\\
110 Shaffer Road, Santa Cruz, CA 95060 USA
}
\date{{\tt manoukisn@niaid.nih.gov \hspace{5mm} Eric.Anderson@noaa.gov}}

\date{{\tt manoukisn@niaid.nih.gov \hspace{5mm} Eric.Anderson@noaa.gov}}

\begin{document}
\twocolumn[
  \maketitle
  \vspace{-5mm}
\begin{onecolabstract}
The computer programs most users interact with daily are driven by a graphical
user interface (GUI). However, many scientific applications are used with a
command line interface (CLI) for the ease of development and increased flexibility this mode
provides. Scientific application developers would benefit from being able to
provide a GUI easily for their CLI programs, thus retaining the advantages
of both modes of interaction.
GuiLiner is a generic, extensible and flexible front-end designed to ``host'' a wide
variety of data analysis or simulation programs. Scientific application
developers who produce a correctly formatted XML file describing their program's options and
some of its documentation can immediately use GuiLiner to produce a carefully
implemented GUI for their analysis or simulation programs.

\end{onecolabstract}
\paragraph*{Key-Words}:
Graphical user interfaces, XML, Computer applications, Software interfaces
\vspace{5mm}
]
\saythanks


\renewcommand{\thefootnote}{\alph{footnote}}
\section{Introduction}
Computer applications for scientific research generally receive user input through a
command line interface (CLI) or through a graphical user interface (GUI). Each 
has advantages and shortcomings. For example, 
GUIs provide immediate accessibility and a familiar mode of interaction for
most users. On the other hand, the CLI allows for batch processing, inclusion of
the program in shell scripts, and the
retention of execution parameters. CLI programs also require less development
time and are more portable across different computer operating systems.

In this paper we present guiLiner, an application designed to bridge the gap
between CLI and GUI modes for computer programs used in scientific research. GuiLiner
is a generic, extensible and flexible front-end designed to ``host'' a wide
variety of data analysis or simulation applications. It is geared primarily toward
the scientific application development community, which can realize several
unique benefits from its use, beginning with the elimination of time
spent writing code to generate a GUI. 

The task of creating a generic GUI for biological scientific applications is made simpler
by the fact that most of them follow a simple interaction model: 1) 
the user provides data and parameters to the application 2) the
algorithm is executed on these and 3) the results of the analysis are returned. Each of
these steps is generally atomic. 

Since step 1) can involve many options, it is here that CLI based applications
can become cumbersome to use or may be intimidating to inexperienced users.
GuiLiner focuses on this step and on step 2).  Without modifying the original
CLI program, guiLiner provides a way for users to quickly see the available
program options, read documentation and set the value of each option, and then
execute the program, all from within a familiar ``point-and-click'' environment.

\section{Implementation}

GuiLiner is written in the
Java programing language, and virtual machines capable of executing it are
available for current versions of Microsoft Windows, Mac OS X and many types of
UNIX based operating systems such as GNU/Linux (for a full list, please see \\
http://www.java.com:80/en/download/manual.jsp).
GuiLiner operates by parsing an XML configuration file which contains
information on the CLI-based application being hosted, its options,
documentation and some details about how guiLiner itself should display this
information (Figure 1). This scheme allows almost unlimited extensibility, so
that the feature-set of guiLiner can be increased with later releases.

\begin{figure}[!tpb]
\centering
\includegraphics[width=80mm]{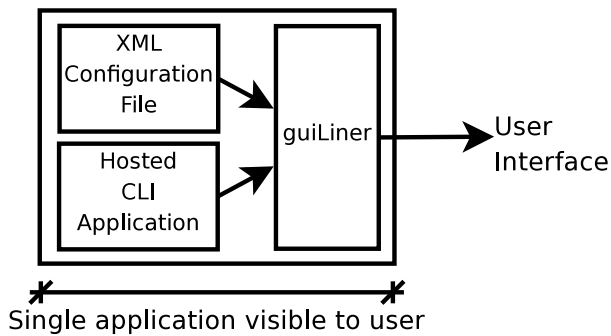}
\caption{Schematic diagram showing how GuiLiner, the hosted analysis application and 
the XML confiuration file are used together to present a single GUI-driven
application to the user.}\label{fig1}
\end{figure}

The GUI provided by guiLiner features a rapid
visual summary of which options are required, set, or unset in a color-coded
option tree;  integrated display of documentation specific to each program
option; facilities for saving the values of options used for a particular
execution of the hosted application;   and the ability to view and save to
disk program run-time output and/or errors.

In addition to the rapid display of selected and required options, efficient
option information retrieval and runtime results, guiLiner's layout is designed
to put commonly used functions within easy each. 
Besides the usual menu bar (Figure 2 \#1) containing a 
custom help browser and XML save-open options among other settings, 
there is also a button bar for functions commonly used during option setting
(Figure 2 \#2). These include functions to preview the command line, rest all
options, manipulate the option tree and run the hosted program. Use of these is
described in more detail in the next section.

\begin{figure}[!tpb]
\centering
\includegraphics[width=80mm]{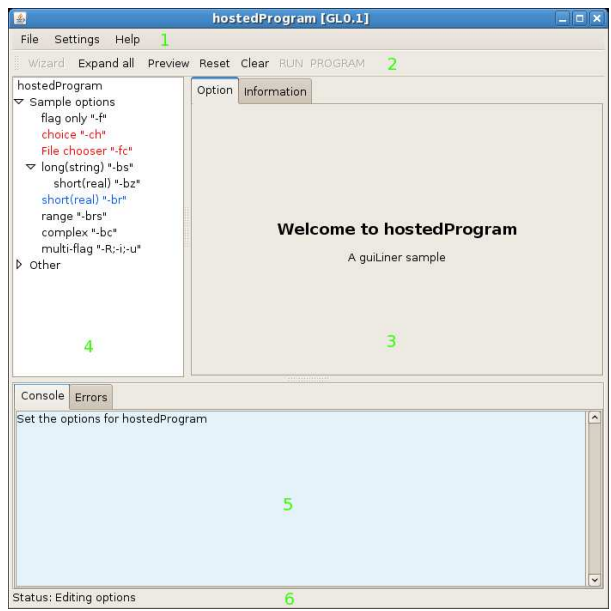}
\caption{GuiLiner hosting a sample program, in Editing mode.}\label{overview}
\end{figure}

We have found few other efforts to create a generic user interface. Some of
these are not focused on the scientific computing community, and so aim to
accommodate a wider variety of CLI programs. These usually take the form of
widget sets that can be configured to create a GUI. While this approach is more
flexible it is usually also more time consuming and less extensible. Other
generic GUI programs use a ``Wizard'' interface, which is both flexible and easy
to deploy, but lack the visual summary and interactivity that guiLiner offers.
The
advantage of guiLiner over either of these approaches is that it is designed to represent a
single mode of interaction that is common in scientific computing, which allows
it to be employed very quickly and at the same time makes it very effective for
hosting these types of programs. To date we have not found
any other applications which fill this exact niche.

\section{Usage Overview}
Most user interaction with guiLiner involves selecting options from the option tree 
(Figure 2 \#4) by clicking on them. When selected, an
interface to manipulate
that option is displayed in the options pane (Figure 2 \#3). The exact
interface will vary
depending on which type of option is being displayed: it may have a text box where
a value can be input, a set of buttons that allows the user to set the option to one of
several given values, or a dialog box for navigating the file directory to 
find an input file, etc. In all cases the user is able to get more information on
the option by clicking the ``Information'' tab in the option pane. 

The colors in the option tree give a quick visual guide to the run
settings. Red = option
is required, value not yet specified; Black = option not required, value not yet
specified; Blue = a value for the option has been specified and will be used for
program execution.

Clicking the Preview button (Figure. 2 \#2) causes the command line
to be assembled from the values currently specified by the user and prints it to
the console panel (Figure. 2 \#5).  This is particularly useful for
``transitional'' users who are gaining familiarity with the command line
environment, but are not yet fully comfortable with it.  Using this facility
then saving the console contents is an easy way to save run settings. An
alternative method it to save the entire XML file with the selected options
already set. Though this is more cumbersome to read, it does allow guiLiner to
automatically load the settings used in that particular execution.

When the RUN PROGRAM button (Figure. 2 \#2) is clicked, guiLiner
uses a system call to execute the CLI program with the options assembled by the
user in guiLiner.  Program output to stdout goes to the console panel from where
it may be viewed or saved to disk as a text file.  Program output to stderr is
directed to the Errors panel and the user is notified of errors in the status
bar (Figure. 2 \#6).  Any program output directed to files goes
to those files specified either by an absolute path or by a path relative to
the current working directory (exactly as if the program were run from the CLI).
 guiLiner is not designed for interactive display of program {\em output},
 though future versions could allow simple GUI-driven output display using
 developer provided scripts and, for example, the R statistical computing
 environment \cite{R2007}.  


For ease of distribution and installation to end-user machines guiLiner, the XML
configuration file and the CLI executable can be distributed as an installer.
There are several excellent installer platforms available which could streamline
this process, such as the platform independent IzPack (available at 
http://izpack.org/). 

Details on the XML file specification, option types, the application
executable and source code, and discussion forums are available at
http://guiliner.sourceforge.net. Also at this web site there are
sample XML configuration files for a variety of bioinformatic and population
genetic analysis programs including Exonerate \cite{Slater2005}, 
IM \cite{Hey2004}, Makesamples \cite{Hudson2002} and Spip \cite{Anderson2005}.
A Document Type Definition (DTD) file is distributed with guiLiner to automate
XML configuration file generation and to allow error checking.

We also distribute
there a small C library for C or C++ programmers that simplifies command-line
parsing and error checking, and allows the documentation for each option to be
written and stored in the source code.  This documentation may be printed by
the program in short-help format, long-help format, UNIX man page format, and
guiLiner XML format.  The guiLiner XML format can be read directly by guiLiner
so that any updates to the program can be immediately translated to the guiLiner
GUI.

We encourage contributions to the source code or comments on guiLiner.

\section{Conclusions}
  GuiLiner is an effective ``wrapper'' for a wide variety of biological analysis
  and simulation software. Application developers will be able to offer a
  functional and carefully implemented GUI to their CLI-driven software with
  little effort. At the same time, guiLiner should make a wider variety of 
  applications immediately available for the analyses of researchers who are not
  familiar with the CLI or are beginning to learn about it.

\section{Acknowledgements} 
This research was supported in part by the Intramural Research Program of the 
NIH, NIAID. The authors would like to thank J. Hey, R. Hudson and J Garza for graciously
permitting us to present their programs bundled with guiLiner to serve as examples.

\end{document}